\documentclass[aps,pre,twocolumn,showpacs]{revtex4}

\usepackage{graphicx,epsfig,subfigure}
\usepackage{amsmath,amssymb}

\begin{document}

\title{Universal interpretation of efficacy parameter in perturbed nonequilibrium systems}
\author{Sourabh Lahiri}
\email{lahiri@iopb.res.in}

\author{A. M. Jayannavar}
\email{jayan@iopb.res.in}

\affiliation{\vspace{0.5cm}
Institute of Physics, Sachivalaya Marg, Bhubaneswar - 751005, India}


\newcommand{\nwc}{\newcommand}
\nwc{\la}{\langle}
\nwc{\ra}{\rangle}
\nwc{\lw}{\linewidth}
\nwc{\nn}{\nonumber}
\nwc{\Ra}{\Rightarrow}
\nwc{\dg}{\dagger}

\nwc{\Tr}[1]{\underset{#1}{\mbox{\large Tr}}~}
\nwc{\pd}[2]{\frac{\partial #1}{\partial #2}}
\nwc{\ppd}[2]{\frac{\partial^2 #1}{\partial #2^2}}

\nwc{\zprl}[3]{Phys. Rev. Lett. ~{\bf #1},~#2~(#3)}
\nwc{\zpre}[3]{Phys. Rev. E ~{\bf #1},~#2~(#3)}
\nwc{\zpra}[3]{Phys. Rev. A ~{\bf #1},~#2~(#3)}
\nwc{\zjsm}[2]{J. Stat. Mech. ~#1~(#2)}
\nwc{\zepjb}[3]{Eur. Phys. J. B ~{\bf #1},~#2~(#3)}
\nwc{\zrmp}[3]{Rev. Mod. Phys. ~{\bf #1},~#2~(#3)}
\nwc{\zepl}[3]{Europhys. Lett. ~{\bf #1},~#2~(#3)}
\nwc{\zjsp}[3]{J. Stat. Phys. ~{\bf #1},~#2~(#3)}
\nwc{\zptps}[3]{Prog. Theor. Phys. Suppl. ~{\bf #1},~#2~(#3)}
\nwc{\zpt}[3]{Physics Today ~{\bf #1},~#2~(#3)}
\nwc{\zap}[3]{Adv. Phys. ~{\bf #1},~#2~(#3)}
\nwc{\zjpcm}[3]{J. Phys. Condens. Matter ~{\bf #1},~#2~(#3)}
\nwc{\zjpa}[3]{J. Phys. A: Math. Theor. ~{\bf #1},~#2~(#3)}
\nwc{\zpjp}[3]{Pram. J. Phys. ~{\bf #1},~#2~(#3)}
\nwc{\zpa}[3]{Physica A ~{\bf #1},~#2~(#3)}

\begin{abstract}

    The fluctuation theorems have remained one of the cornerstones in the study of systems that are driven far out of equilibrium, and they provide strong constraints on the fraction of trajectories that behave atypically in light of the second law. They have mainly been derived for a predetermined external drive applied to the system. However, to improve the efficiency of a process, one needs to incorporate protocols that are modified by receiving feedbacks about the recent state of the system, during its evolution. In such a case, the forms of the conventional fluctuation theorems get modified, the correction term involving terms that depend on the way the reverse/conjugate process is defined, namely, the rules of using feedback in order to generate the exact time-reversed/conjugate protocols. We show in this paper that this can be done in a large number of ways, and in each case we would get a different expression for the correction terms. This would in turn lead to several lower bounds on the mean work that must be performed on the system, or on the entropy changes. Here we analyze a form of the extended fluctuation theorems that involves the efficacy parameter, and find that this form gives rise to a  lower bound for the mean work that retains a consistent physical meaning regardless of the design of feedback along the conjugate process, as opposed to the case of the previously mentioned form of the modified fluctuation theorems.

\end{abstract}

\maketitle{}

\section{Introduction}

The theory of classical thermodynamics has been well supported by the methods of equilibrium statistical physics, which in turn is based on the Boltzmann probability distribution for the states of a system in phase space. For systems that are slightly away from equilibrium (where the perturbation to the system can be assumed to be linearly coupled to the applied weak force field), the linear response theory characterizes the response of the system in terms of equilibrium correlation functions. However, there are only very few exact results that remain valid even when the system is driven arbitrarily out of equilibrium.
The fluctuation theorems have been one of them. First introduced for thermostatted systems following dissipative but deterministic dynamics by Evans and Searles \cite{eva93,eva94}{}, it has produced several closely related equalities \cite{jar97,jar97a,cro98,cro99,sei05,sei08}, some theorems holding in general for all systems while the others being satisfied for special cases (for example, in the limit of a large observation time). In a nutshell, these theorems have the following general form:
\begin{equation}
  \frac{P(X_t)}{P^T( X_t^T)} = e^{X_t},
\end{equation}
where $X_t$ is some observable that is to be measured and is in general a path function, $P(X_t)$ is the probability density of this observable along a process (usually parametrized by an externally controlled time-dependent protocol $\lambda(t)$) and $ P^T$ is the probability density whose functional form is related to $P$ through a conjugate transformation which is not necessarily  time-reversal. Its argument $X_t^T$ is the value assumed by the observable along the transformed trajectory in phase space. Some observables that have been shown to follow the fluctuation theorems are entropy, work and heat \cite{jar97,jar97a,cro98,cro99,sei05,sei08,han11,sei12}. It is easy to see that the theorems are closely related to the second law, in the sense that the averaged observable $\la X_t\ra$ always abides by the inequalities dictated by the second law. For instance, the total entropy of the system plus environment can be shown to give rise to the inequality \cite{sei05,sei08}
\begin{equation}
  \la\Delta s_{tot}\ra \ge 0.
\end{equation}
What this inequality says is that, even though some individual realizations of the experiment may observe a decrease in the total entropy, its \emph{average} value must never decrease with time.

An issue that has recently attracted much interest is the use of feedback-controlled protocol to enhance the efficiency of such a process \cite{sag08,sag10,sag12,hor10,pon10,lah12,ran12,ran12a,sei12a,cao09}. Instead of using a predetermined protocol, we can perform intermediate measurements on some observable and modify our protocol accordingly. We are in general not interested in the measuring device involved in the experiment (which definitely belongs to the ``universe'' in which second law must be obeyed). It is obvious that the form of the second law, now dealing with only a part of the universe under consideration, needs to be modified. As expected, the correction term involves the properties of the ignored subpart, in the form of the mutual information gained through measurements and gives rise to the following inequality:
\begin{equation}
  \la X_t\ra \ge -\la I\ra.
  \label{law2_mod}
\end{equation}
Here, the mutual information is defined as  \cite{sag08,sag10,sag12,hor10,pon10,lah12}
\begin{equation}
  I \equiv \ln \frac{p(m_1|x_1)p(m_2|x_2)\cdots P(m_N|x_N)}{P(m_1,m_2,\cdots,m_N)},
  \label{I}
\end{equation}
where the presence of measurement errors is assumed. In the arguments of the conditional probability $p(m_i|x_i)$, $m_i$ is the outcome when a measurement is performed, while $x_i$ is the actual value of the observable. The averaging has been done over all possible phase space trajectories {\it as well as} over all possible protocol functions, which results due to the different feedback rules corresponding to different outcomes. Since $\la I\ra$ is a relative entropy and consequently always non-negative \cite{cov}, the above inequality provides us with the privilege of extracting work from or obtaining a decrease in entropy of the system, provided we can  formulate a nice feedback algorithm. Note that eq. (\ref{law2_mod}) only says that the lower bound for the dissipated work (or entropy) \emph{can} be decreased by an amount proportional to the average mutual information. However, it does not provide any clue to what a good feedback algorithm should be. In this sense, it is reasonable to look for a quantity that actually provides a quantitative measure of how efficient an employed feedback procedure is. This is the so-called {\it efficacy parameter} \cite{sag12} defined below.
In this paper, we clarify the physical meaning and some of the  properties of this parameter. The Jarzynski equality \cite{jar97,jar97a} has two different generalized forms in the presence of feedback.  The more commonly used form of the extended Jarzynski Equality (EJE) is  \cite{sag08,sag10,sag12,hor10,pon10,lah12}
\begin{equation}
  \la e^{-\beta W_d[X] - I[X,M]}\ra = 1.
  \label{EJEa}
\end{equation}
$W_d$ is the dissipated work defined through $W_d = W-\Delta F$, where $\Delta F$ is the difference between free energies at the end and at the beginning of a particular protocol.
Here, the reverse trajectories are generated by simply reversing the sequence of one of the forward protocols and we have used upper case letters to denote the full path of successive phase points or measured outcomes: $X\equiv (x_0,x_1,\cdots,x_\tau)$, $M\equiv (m_1,m_2,\cdots, m_N)$. The time has been discretized as $(t_0,t_1,t_2,\cdots,t_N,t_{N+1})$. We will set $t_0=0$ and $t_{N+1}=\tau$. As we will show in this paper, $I[X,M]$ will in general be replaced by a different physical quantity $\phi[X,M]$,  if we choose to use feedback along the reverse process as well, as will be detailed below. 

Other than eq. (\ref{EJEa}), there is yet another form of the extended Jarzynski Equality that has been introduced in the literature \cite{sag10,sag12}:
\begin{equation}
  \la e^{-\beta W_d[X]}\ra = \gamma.
  \label{EJEb}
\end{equation}
The efficacy parameter $\gamma$ is a functional of the feedback control used along the forward process, and that determines the extent to which 
the feedback is efficient (i.e., more work can be extracted from the system). If $\gamma=1$, then we would have the Jarzynski equality in absence of feedback. The efficacy parameter has been measured experimentally \cite{sag10a} and equations (\ref{EJEa}) and (\ref{EJEb}) have been verified. Using the Jensen's inequality, we have the dissipated work bounded from below through the relation
\begin{equation}
  \la W_d[X]\ra \ge -\ln\gamma
\end{equation}
Although this inequality looks similar to the one stated before, it may seem to be a trivial statement because it is a simple consequence of the definition of $\gamma$. However, the fact that this definition lends  a very clear physical meaning to the efficacy parameter and that this meaning can be exploited to experimentally measure $\gamma$ (without using the definition eq. (\ref{EJEb})) underlines the importance of the above bound. We explicitly show that $\gamma$ retains the simple meaning when we extend eq. (\ref{EJEb}) to driven systems that make excursions among nonequilibrium steady states under arbitrary feedback-controlled protocols, whereas the physical meaning of $\phi[X,M]$ depends on the design of feedback along the reverse process.

\section{Extended Jarzynski Equality}
\label{sec:EJE}

\subsection{Blind time-reversal of protocol} \label{case:I}

  As shown in \cite{sag08,sag10,sag12,hor10,pon10,lah12}, in presence of information gain and feedback applied \emph{only} along the forward trajectory, the Jarzynski Equality gets modified to eq. (\ref{EJEa}).
  This relation is easily derived from  the detailed fluctuation theorem, in the case when  the reverse protocol is the blind time-reversal of the corresponding protocol along the forward process. Then the ratio of the forward and reverse trajectories become \cite{hor10,pon10,lah12}
  \begin{align}
    \frac{P[X,M]}{\tilde P[\tilde X;M]} &= \frac{P[X]}{\tilde P[\tilde X]} \times  \frac{p(m_1|x_1)p(m_2|x_2)\cdots P(m_N|x_N)}{P(m_1,m_2,\cdots,m_N)}, \nn\\
    &= e^{\beta W_d[X]+I[X,M]}.
    \label{EJE1}
  \end{align}
In the above equation, $\tilde X$ is the time-reversed trajectory $(\tilde x_\tau,\tilde x_N, \cdots, \tilde x_0)$, and $\tilde P[\tilde X]$ is the probability density for this time-reversed trajectory (for simplicity of notation, we would always use the forward time $t$ even during the reverse process, in place of using the actual time $\tilde t = \tau-t$ elapsed during the reverse process).  If the state is denoted only by the position variable, then we have $\tilde x_i=x_i$. $\tilde P[\tilde X;M]$ is the probability path functional for the \emph{actual reverse process} being carried out. This is different from $\tilde P[\tilde X,\tilde M]$ which gives the probability of obtaining the reverse phase space trajectory along with the reversed measured outcomes, provided measurements are performed at the same time instants as in the forward process. The expression for the latter quantity has no explicit reference to the actual feedback procedure followed to generate the reverse process (i.e., its physical meaning does not change due to change in feedback procedure for reverse process). A simple cross-multiplication followed by integration over $X$ and $M$ will give rise to the  modified Jarzynski Equality, eq. (\ref{EJEa}).

\subsection{Using feedback to generate reverse process} 

\label{case:II} The reverse process can also be defined by designing a suitable feedback procedure to generate the time-reversed protocol, which does not violate causality \cite{anu12}. The feedback procedure is as follows: we first measure system observable at time $t=\tau$, and if the measurement outcome is equal to the forward outcome at time $t=t_N$, then we drive the system using the reverse protocol $\lambda(m_{t_N})$ (note that the  time instants for measurement along the backward process are not the same as the forward ones, but are shifted). Continuing with this process, i.e., changing the protocol according to measurement outcome $m(t_{k-1})$ at time $t=t_k$, the reverse protocol is exactly reproduced without ever violating causality. This the new relation
  \begin{equation}
    \frac{P[X,M]}{\tilde P[\tilde X;M]} = e^{\beta W_d[X]+\Delta s_p[X,M]},
    \label{EJE2}
  \end{equation}
  where 
  \begin{equation}
    \Delta s_p[X,M] \equiv \ln\frac{p(m_0|x_0)p(m_1|x_1)\cdots p(m_{N}|x_{N})}{p(m_N|\tilde x_{N+1})p(m_{N-1}|\tilde x_{N})\cdots p(m_0|\tilde x_1)}.
  \end{equation}
  Here, $\Delta s_p$ represents a disorder parameter, but is not the mutual information as defined in \cite{sag10}.
  Of course, equations (\ref{EJE1}) and (\ref{EJE2}) contain different information as well as provide different bounds for the average total entropy change for the system.

\subsection{The most general case}

\begin{figure}[!h]
\centering
\includegraphics[width=0.8\linewidth]{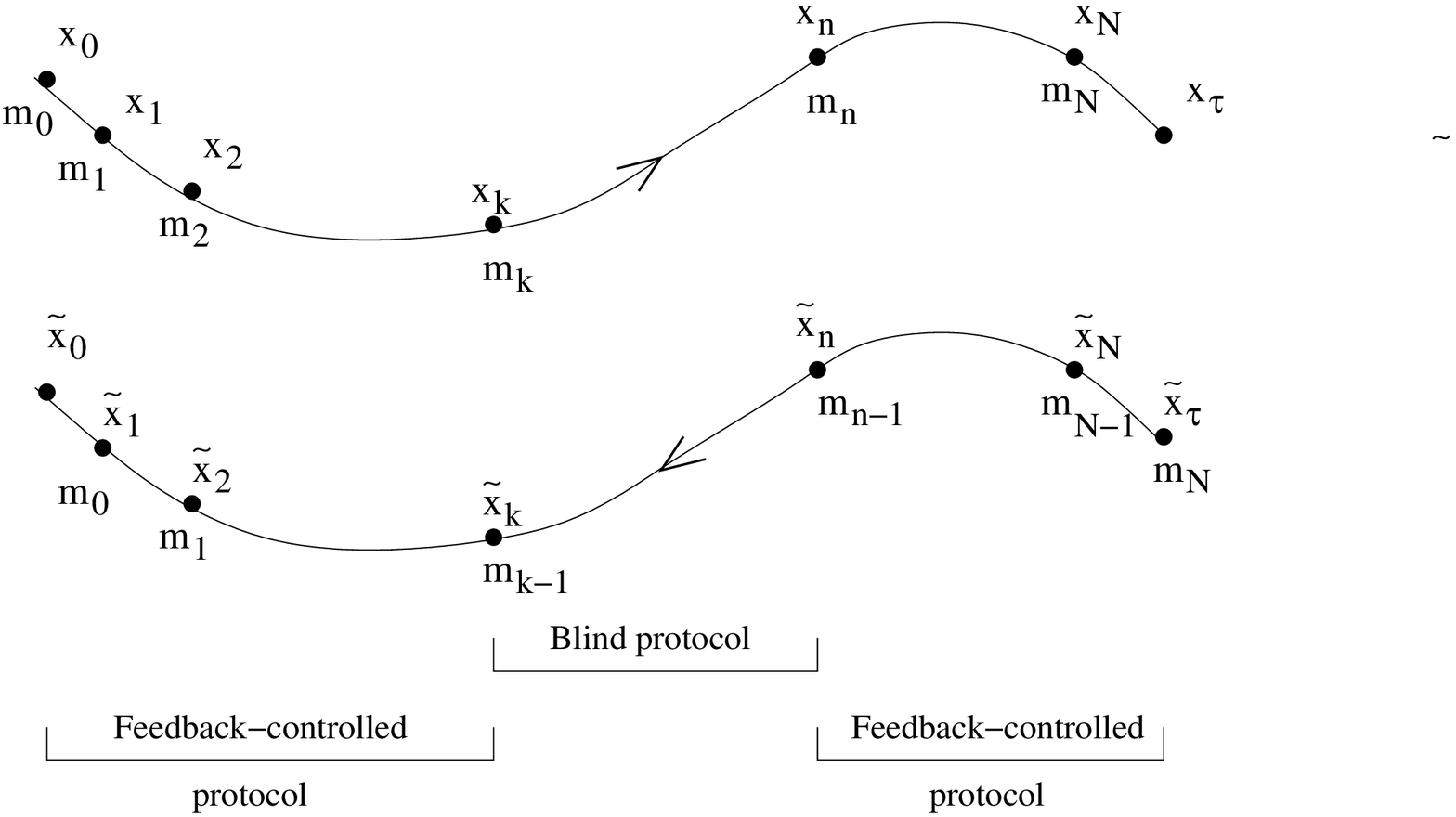}
\caption{The figure shows the representative plot of the phase pace trajectory in a simple case for the most general form of feedback that can be considered during the reverse process (the lower curve) corresponding to the forward process (the upper curve). First, from $t=\tau$ to $t=t_n$, we have the feedback-controlled reverse protocol, while from $t=t_{n}$ to $t=t_k$ we have used the blindly time-reversed protocol. In the final part, up to $t=0$, we revert to the feedback-controlled protocol.}
\label{fig1}
\end{figure}

\label{case:III}

The general form of the probability density for a forward trajectory in presence of feedback is given by \cite{lah12}
\begin{align}
P[X,M] =& p(m_0|x_0)P_{m_0}[x_0\to x_1]p(m_1|x_1)P_{m_1}[x_1\to x_2] \nn\\
&\cdots p(m_N|x_N)P_{m_N}[x_N\to x_\tau].
\end{align}

The general protocol to generate the reverse process would be to use both the protocols \ref{case:I} and \ref{case:II} at random during the reverse process. Let us take one simple case when, along the reverse process, up to time $t=\tau-t_n$, we use the feedback-controlled protocol, then from time $t=t_n$ to $t=t_k$, we use the blindly applied reverse protocol, and finally from $t=t_k$ to $t=0$, we once again use the feedback-controlled reverse protocol (see figure \ref{fig1}).  Here, $k$ and $n$ can be any two integers chosen at random from the set $\{1,2,\cdots,N\}$. In this case, the reverse process becomes,
  \begin{align}
    &\tilde P[\tilde X;M] = p(\tilde x_\tau) p(m_N|\tilde x_\tau) P_{m_N}[\tilde x_\tau\to \tilde x_{N}] \nn\\
    &\hspace{1cm} p(m_{N-1}|\tilde x_N) P_{m_{N-1}}[\tilde x_N\to \tilde x_{N-1}] \nn\\
    &\cdots\hspace{0.5cm}  p(m_{n}|\tilde x_{n+1}) P_{m_n}[\tilde x_{n+1}\to \tilde x_n] \nn\\
    &\hspace{1cm} p(m_{n-1},\cdots,m_k) P[\tilde x_n \to \tilde x_k] p(m_{k-1}|\tilde x_k) \nn\\
    &\hspace{1cm} P[\tilde x_k \to \tilde x_{k-1}] \cdots p(m_0|\tilde x_1) P[\tilde x_1\to \tilde x_0].
    \label{genrev}
  \end{align}
  Therefore, we arrive at
  \begin{align}
    \frac{P[X,M]}{\tilde P[\tilde X;M]} =& \frac{P[X]}{\tilde P[\tilde X]}~ \frac{p(m_k|x_k)\cdots p(m_{n-1}|x_{n-1})}{p(m_{k},\cdots,m_{n-1})} \nn\\
    &\times \frac{p(m_0|x_0)\cdots p(m_{k-1}|x_{k-1})}{p(m_0|\tilde x_1)\cdots p(m_{k-1}|\tilde x_k)} \nn\\
    &\times \frac{p(m_n|x_n)\cdots p(m_N|x_N)}{p(m_n|\tilde x_{n+1})\cdots p(m_N|\tilde x_\tau)} \nn\\
    =& \exp\left[{\beta W_d+I^1 + \Delta s_p^1 + \Delta s_p^2}\right],
    \label{EJE3}
  \end{align}
  where
  \begin{align}
    I^1 &= \ln\frac{p(m_k|x_k)\cdots p(m_{n-1}|x_{n-1})}{p(m_{k},\cdots,m_{n-1})}; \label{I1} \\
    \Delta s_p^1 &=\ln \frac{p(m_0|x_0)\cdots p(m_{k-1}|x_{k-1})}{p(m_0|\tilde x_1)\cdots p(m_{k-1}|\tilde x_k)}; \label{sp1}\\
    \Delta s_p^2 &= \ln\frac{p(m_n|x_n)\cdots p(m_N|x_N)}{p(m_n|\tilde x_{n+1})\cdots p(m_N|\tilde x_\tau)}. \label{sp2}
  \end{align}

  We are thus led to  a different extended detailed fluctuation theorem where disorder parameters $\Delta s_p^1$, $I^1$ and $\Delta s_p^2$ are different as they contain different information about the feedback process. Moreover, these are not unique and may depend on different reverse protocols not considered here. Thus, it is clear that the $\phi[X,M]$ in eq. (\ref{EJE1}) does not have a unique interpretation and clearly depends  on the manner of feedback along the backward process. From eq. (\ref{EJE3}), one can obtain the related integral fluctuation theorem, namely 
\begin{equation}
\la \exp[-\beta W_d- I^1 - \Delta s_p^1 - \Delta s_p^2]\ra = 1
\end{equation}

All the three results derived above, eqs. (\ref{EJE1}), (\ref{EJE2}) and (\ref{EJE3}), can be written in compact form as
\begin{equation}
\frac{P[X,M]}{\tilde P[\tilde X;M]} = e^{\beta W_d+\phi[X,M]},
\label{EJE_gen}
\end{equation}
where $\phi[X,M]$ is a functional of the phase space space trajectory as well as of the measurement trajectory (i.e., sequence of measured outcomes).

 We next turn our attention to the efficacy parameter.

\section{Efficacy parameter in presence of general feedback}

\label{gamma_JE}

In absence of feedback, the Jarzynski equality is given by \cite{jar97,jar97a,cro98,cro99}
\begin{equation}
\la e^{-\beta W_d[X]}\ra = 1.
\end{equation}
In presence of feedback, the right hand side of the above relation will in general not be unity.
The efficacy parameter for the feedback in this case (when system is initially at thermal equilibrium with the medium) is defined as
\begin{equation}
\gamma = \la e^{-\beta W_d[X]}\ra.
\label{eff}
\end{equation}
It can be readily seen that lesser the amount of dissipated work, more is the magnitude of the efficacy parameter, as is desirable for a quantity that decides the efficiency of feedback. Now we use the most general case for obtaining the reverse trajectories, namely the case \ref{case:III}.

Therefore, we get
\begin{align}
\gamma &= \int \mathcal{D}[X,M] P[X,M] e^{-W_d[X,M]} \nn\\
    &=  \int \mathcal{D}[X,M] \tilde P[\tilde X;M]e^{\Delta s_p^1 + I^1 + \Delta s_p^2} \nn\\
    &= \int \mathcal{D}[X,M] \tilde P[\tilde X] p(m_0|x_0) p(m_1|x_1) \nn\\
    & \hspace{4cm}\cdots p(m_N|x_N) \nn\\
    & = \int \mathcal{D}[\tilde X,\tilde M] \tilde P[\tilde X] p(\tilde m_0|\tilde x_0) \cdots p(\tilde m_N|\tilde x_N) \nn\\
    &= \int \mathcal{D}[\tilde X,\tilde M] \tilde P[\tilde X,\tilde M] 
    = \int \mathcal{D}[\tilde M] \tilde P[\tilde M],
\label{eff_JE}
\end{align}
  where we have used the detailed fluctuation theorem (\ref{EJE_gen}), and the definitions (\ref{sp1}), (\ref{I1}) and (\ref{sp2}). The time-reversibility of measurements has been assumed: $p(m_i|x_i) = \tilde p(\tilde m_i|\tilde x_i)$ \cite{sag10}. Once again we may recall that the two quantities $\tilde P[\tilde X;M]$ and $\tilde P[\tilde X,\tilde M]$, both of which appear in the above derivation, are different. 
In more general cases, the steps 2 and  3  can be written as
\begin{align}
\tilde P[\tilde X;M]& e^{\phi[X,M]} = \tilde P[\tilde X;M] \exp\left(\sum_i\Delta s_p^i + \sum_j I^j\right) \nn\\
&= \tilde P[\tilde X] p(m_1|x_1) \cdots p(m_N|x_N) \nn\\
&= \tilde P[\tilde X] \tilde p(\tilde m_1|\tilde x_1) \cdots \tilde p(\tilde  m_N|\tilde x_N) \nn\\
&= \tilde P[\tilde X,\tilde M].
\end{align}
Here, the summations $\sum_i$ and $\sum_j$ run over all the time intervals in which the reverse protocols have been executed using feedback and by blind time-reversal, respectively. We find that although the form of $\tilde P[\tilde X;M]$ contains detailed information about the actual feedback procedure used along the backward process, when it is multiplied by the factor $e^{\phi[X,M]}$, we obtain $\tilde P[\tilde X,\tilde M]$  whose form does not contain any such information. This is the reason behind the fact that the efficacy parameter retains the same physical meaning in each case, namely, it is  the \emph{total probability to observe the time-reversed outcomes for the  measurements performed along the reverse process} \cite{sag08,sag10,sag12,lah12}.

 This  derivation of $\gamma$ for the most general case is our central result. It simply shows that it retains the same meaning as for the specific cases considered in the earlier literature \cite{sag10,sag12,hor10,pon10,lah12,ran12,ran12a,sei12a}.

\section{The three detailed fluctuation theorems}

We will be generalizing our treatment to the other detailed fluctuation theorems, which involve the non-adiabatic entropy production (change in entropy due to shift between two different steady states) and adiabatic entropy production (change in entropy caused due to maintenance of a given steady state). 
The total entropy change by definition is the sum of entropy changes in the system ($\Delta s$) and in the medium ($\Delta s_m$): $\Delta s_{tot}=\Delta s+\Delta s_m$. 
Recently it has been observed that while generalizing the second law for systems making transitions between steady states, the total entropy production can also be split into two distinct parts such that each part, interestingly, follows a detailed fluctuation theorem \cite{esp10,esp10a,esp10b,gar12}: 
\begin{equation}
\Delta s_{tot} = \Delta s_{na} + \Delta s_a.
\end{equation}
The averages of all these three quantities are always non-negative, thereby providing a new twist to the second law. $\Delta s_a$ is related to the housekeeping heat $Q_{hk}$, while $\Delta s_{na}$ is the sum of the entropy change of the system and the entropy produced due to excess heat $Q_{ex}$  \cite{esp10,esp10a,esp10b,gar12}.

In the case of adiabatic and non-adiabatic entropy productions, the concept of dual dynamics is very helpful. Under the dual dynamics, if the system is allowed to reach the corresponding steady state, then the steady-state distribution $\rho_{SS}$ retains the same form, but the probability current reverses its sign \cite{gar12,jar06}. Hatano and Sasa had shown that the physical meaning of the nonadiabatic entropy becomes clear in the dual dynamics formalism \cite{hat01}. These detailed fluctuation theorems are taken up in the following discussion.

\subsection{Total entropy} 

When the initial state of the system is not at thermal equilibrium with the bath, then in absence of feedback, the following ratio is obtained between the forward and the reverse trajectories \cite{sei05,sei08,sei12}:
\begin{equation}
\frac{P[X]}{\tilde P[\tilde X]} = e^{\Delta s_{tot}[X]},
\end{equation}
from which the following integral fluctuation theorem can be obtained:
\begin{equation}
\la e^{-\Delta s_{tot}[X]}\ra = 1.
\end{equation}
In presence of feedback, the right hand side will in general be different from unity. For this general case, instead of eq. (\ref{EJE_gen}), we would get the following ratio between the forward and reverse paths:
\begin{equation}
\frac{P[X,M]}{\tilde P[\tilde X;M]} = e^{\Delta s_{tot}[X]+\phi[X,M]}.
\end{equation}
We now consider the case with general reverse protocol. We define the efficacy parameter as
  \begin{align}
    \gamma_{tot} = \la e^{-\Delta s_{tot}[X]}\ra 
  \end{align}
Proceeding in exactly the same way as before, we find 
\begin{align}
  \gamma_{tot} &= \int \mathcal{D}[X,M] P[X,M] e^{-\Delta s_{tot}[X,M]} \nn\\
  &=  \int \mathcal{D}[X,M] \tilde P[\tilde X;M] e^{\phi[X,M]}\nn\\
  &= \int \mathcal{D}[\tilde M] \tilde P[\tilde M].
\end{align}
Thus, $\gamma_{tot}$ retains the same physical meaning as $\gamma$ for the Jarzynski equality, although here we do not have the constraint of sampling the initial state of the system from the Boltzmann distribution.

\subsection{Nonadiabatic entropy}

  For transitions between nonequilibrium steady states, we have the following detailed fluctuation theorem in absence of feedback \cite{esp10a,gar12,hat01}:
  \begin{equation}
    \frac{P[X]}{\tilde P^\dagger[\tilde X]} = e^{\Delta s_{na}[X]}.
  \end{equation}
The superscript $\dagger$ implies dual dynamics, which implies that keeping the functional form of the protocol same, we are switching to a dynamics where the steady state current simply changes direction as compared to the original dynamics. The tilde symbol over $P$ implies that the protocol for the forward process has been time-reversed after the system has been allowed to follow the dual dynamics. In other words, $\tilde P^\dagger[\tilde X]$ is the probability density for a trajectory along the process generated, in presence of dual dynamics, by the time-reversed protocol.
Similar to the above cases, in presence of feedback, we  have \cite{lah12,sei12a}
\begin{equation}
\frac{P[X,M]}{\tilde P^\dagger[\tilde X;M]} = e^{\Delta s_{na}[X]+\phi[X,M]},
\end{equation}
where the form of $\phi[X,M]$ depends on the way in which feedback is applied along the reverse trajectory, as given in section 2.
  The efficacy parameter in this case is given by
  \begin{align}
    \gamma_{na} &= \la e^{-\Delta s_{na}[X]}\ra = \int \mathcal{D}[X,M] P[X,M] e^{-\Delta s_{na}[X,M]} \nn\\
    &= \int \mathcal{D}[X,M] \tilde P^\dagger [\tilde X;M] e^{\phi[X,M]} \nn\\
    &= \int \mathcal{D}[X,M] \tilde P^\dagger [\tilde X] p(m_1|x_1)\cdots p(m_N|x_N) \nn\\
    &= \int \mathcal{D}[X,M] \tilde P^\dagger [\tilde X,\tilde M] = \int \mathcal{D}[\tilde M]\tilde P^\dagger [\tilde M].
  \end{align}
In the third step, we have used the algebra that has already been shown in the case of the most general protocol (section \ref{gamma_JE}) for extended Jarzynski equality, which results in
\begin{align}
\tilde P^\dagger[\tilde X;M]& e^{\phi[X,M]} = \tilde P^\dagger[\tilde X] \exp\left(\sum_i\Delta s_p^i + \sum_j I^j\right) \nn\\
&= \tilde P^\dagger[\tilde X] p(m_1|x_1) \cdots p(m_N|x_N) \nn\\
&= \tilde P^\dagger[\tilde X] \tilde p^\dagger(\tilde m_1|\tilde x_1) \cdots \tilde p^\dagger(\tilde  m_N|\tilde x_N) \nn\\
&= \tilde P^\dagger[\tilde X,\tilde M].
\end{align}
We have assumed that the measurement errors do not change on changing the dynamics, which is quite reasonable assumption. 
Thus, $\gamma_{na}$ is the net probability for obtaining the time-reversed outcomes along the time-reversed dual dynamics (i.e., the process in which the protocol corresponding to the dual dynamics is time-reversed).

\subsection{Adiabatic entropy}

 The DFT for adiabatic entropy production is given by \cite{esp10a,gar12} 
  \begin{equation}
    \frac{P[X]}{P^\dagger[X]} = e^{\Delta s_{a}[X]}.
  \end{equation}
$P^\dagger[X]$ is the probability density for the path followed by the system in phase space, when the system is evolving under the  dual dynamics.
In presence of feedback, we then have,
\begin{equation}
    \frac{P[X,M]}{P^\dagger[X;M]} = e^{\Delta s_{a}[X]+\phi[X,M]}.
\end{equation}
Since there is no time-reversal involved, the denominator can only consist of the following options:

(1) The same feedback procedure is used to generate the forward process in the dual dynamics as well, in which case we have $\phi[X,M]=0$ (since the error probabilities in the numerator cancel with those in the denominator). Once again, measurement errors are assumed to be independent of the dynamics followed by the system.

(2) One of the forward protocols in the original dynamics is recorded, and this protocol is blindly executed in presence of the dual dynamics, in which case we simply have $\phi[X,M] = I[X,M]$.

(3) We use the above two procedures at random while generating the forward trajectories in presence of dual dynamics, which is the most general case. In this case, however, $\phi[X,M]=\sum_j I^j$, i.e., the summation over $\Delta s_p^i$ will be absent, because the latter quantity never appears in this case.

  The efficacy parameter is 
  \begin{equation}
    \gamma_a \equiv \la e^{-\Delta s_a[X]}\ra, 
  \end{equation}
which  leads to 
\begin{align}
 \gamma_a &= \int \mathcal{D}[X,M] P[X,M] e^{-\Delta s_{a}[X,M]} \nn\\
 &= \int \mathcal{D}[X,M] P^\dagger[X;M] e^{\sum_j I^j} = \int \mathcal{D}[M]P^\dagger[M].
\end{align}

  Therefore, $\gamma_a$ is  the total probability for observing the same outcomes as the initial process with the same protocol, if the system follows the dual dynamics.
  
We thus find that the physical meaning of efficacy parameter can be very generally stated as follows: \emph{it is the total probability to observe the  measured outcomes conjugate to those along the forward protocol, for the intermediate measurements along the process with the corresponding conjugate dynamics.}  Since the efficacy parameters are experimentally measurable, they would provide  more meaningful bounds for the $\la W_d\ra, \la \Delta s_a\ra, \la \Delta s_{na}\ra, \la \Delta s_{tot}\ra$, and these bounds are in fact independent of whether or not feedback is performed along the conjugate process in the actual protocol, but only requires measurements to be performed along the conjugate process. For the other bounds stated in section 2, the values would depend sensitively on whether and how the feedback is performed along the conjugate dynamics, namely, the extended integral fluctuation theorems can be stated as
\begin{equation}
\left<\exp\left(-\Delta s_k - \sum_i\Delta s_p^i - \sum_jI^j\right)\right> = 1
\end{equation}
for $k=1,2,3$ representing $\Delta s_{tot}$, $\Delta s_{na}$ and $\Delta s_a$, respectively, and this would lead to the bounds
\begin{equation}
\la \Delta s_k\ra \ge  - \left<\sum_i\Delta s_p^i\right> - \left<\sum_jI^j\right>.
\end{equation}
As a consequence, arbitrary number of bounds can be computed for this latter case, which is not only confusing, but undermines the motivation behind such calculations. Thus, the efficacy parameter is a far more suitable experimentally measurable quantity that can characterize not only the performance of the system, but can also act as a useful lower bound for the averaged observables.

\vspace{0.5cm}
\section{Conclusions}

In this paper, we have shown that out of the two known forms of the modified fluctuation theorems in presence of feedback, one of the forms is heavily dependent on the way feedback is applied along the conjugate process, and thereby leads to arbitrary number of lower bounds for work done on the system or for the relevant entropy changes (total entropy, nonadiabatic and adiabatic entropies) taking place. On the other hand, the second form, namely the fluctuation theorem expressed in terms of the efficacy parameter, provides a bound for work and entropy changes that carries a clear and consistent physical meaning, irrespective of the manner of application of feedback along the conjugate process. This consistency is robust even when the conjugate process is not the time-reversed process. This study would hopefully help in simpler experimental verification of the extended fluctuation theorems.

\section{Acknowledgement}

One of us (AMJ) thanks DST, India for financial support. We thank Dr. Anupam Kundu for sending us the preprint of his manuscript prior to its publication.

\end{document}